\documentclass[12pt]{article}
\usepackage{amsmath,amssymb,amsthm}
\usepackage{amsfonts,mathrsfs}
\usepackage{hyperref}
\newtheorem{pro}{Proposition}
\newtheorem*{conjecture*}{Conjecture}
\newtheorem*{algorithm*}{Algorithm}

\newtheorem{cor}{Corollary}
\newtheorem{defin}{Definition}

\newtheorem{remark}{Remark}

\newcommand{\diff}[2]{\frac{\partial{#1}}{\partial{#2}}}

\date{January 12, 2023}
\begin{document}

\title{\protect\vspace*{-10mm}Complete characterization of nontrivial local conservation laws\\ and nonexistence of local Hamiltonian structures\\ for generalized Infeld--Rowlands equation}
\author{J. Va\v{s}\'\i\v{c}ek\\[3mm]
  \small Mathematical Institute,
  \\
  \small Silesian University in Opava, Czech Republic
  \\
  \small E-mail: \texttt{jakub.vasicek@math.slu.cz} 
  }
\maketitle

\begin{abstract}
  We characterize all cases when a certain natural generalization of the Infeld--Rowlands equation admits nontrivial local conservation laws of any order, and give explicit form of these conservation laws modulo trivial ones. Furthermore, we prove 
  that the equation under study admits no 
  nontrivial local Hamiltonian and symplectic structures and no nontrivial local Noether and inverse Noether operators; the method of establishing the said nonexistence results can be readily applied to many other PDEs.\looseness=-1 




\end{abstract}



\section{Introduction}\label{in}

In this paper we consider a PDE in one dependent and three independent variables of the form 
\begin{equation}\label{eqn}
u_t=-\left(u_{xxx}+a u_y+f\right)_x\equiv F.
\end{equation}
where $u=u(x,y,t)$, 
$a=\mathrm{const}$, and $f=f(u,u_x)$ is a smooth function of its arguments; the subscripts indicate partial derivatives in the usual manner.

This equation will be hereinafter referred to as 
the {\em generalized Infeld--Rowlands equation}, as it is a natural generalization for the Infeld--Rowlands \cite{ir} equation that arises inter alia in the 
study of the stability of the Ginzburg--Landau equation 
and is recovered from \eqref{eqn} upon setting $a=1$ and $f=u_x^2$.

Note also that \eqref{eqn} with $a=1$ and $f=u^2/2$ can be seen as a weakly two-dimensional generalization of the following form, see e.g.\ \cite{hn} and references therein, of the well-known spatially one-dimensional version of
the Kuramoto--Sivashinky equation
\[
u_t=-u_{xxxx}-u_{xx} - u u_x,
\]
which is recovered from \eqref{eqn} upon setting $a=1$ and $f=u^2/2$ and dropping the $y$-dependence of $u$.


Below we provide inter alia a complete characterization of nontrivial local conservation laws of all orders for (\ref{eqn}). 
Recall that c
onservation laws play an important role in the theory of PDEs, cf.\ e.g.\ \cite{kvv, olv93, rc, r2, s08, t, xw, wz} and references therein. 
For one, it is natural to require that discretizations employed for numerical solving the PDE under study respect the known conservation laws, see e.g.\ \cite{b-m}. Furthermore, conservation laws  can be used in the course of proving stability, existence and uniqueness of certain kinds of solutions, cf.\ e.g.\ \cite{c-e, c-s}, and the same applies to Hamiltonian structures and proving stability, see e.g.\ \cite{hm} and references therein.

The problem of finding {\em all} inequivalent nontrivial local conservation laws for a given PDE is quite difficult, especially in the case of more than two independent variables, and was successfully addressed only for a rather small number of examples, cf.\ e.g.\ \cite{h, h2, i, sev16,vps, va} and references therein.


We are not aware of previous studies on conservation laws for the generalized Infeld--Rowlands equation \eqref{eqn} or its special cases; on the other hand, the Lie point symmetries of the original Infeld--Rowlands 
equation were completely characterized 
in \cite{fw} where it was also shown that the original Infeld--Rowlands equation does not pass the Painlev\'e test and thus is extremely unlikely to be integrable in the sense of soliton theory; see also \cite{v} for the differential invariant algebra associated to the above point symmetry algebra.


In the present paper we provide a complete list of all cases when \eqref{eqn} with a nonlinear $f$ admits nontrivial local conservation laws, and and for all the cases in question list these conservation laws up to the addition of trivial ones.

Furthermore, we show that (\ref{eqn}) admits no nontrivial local cosymmetries other than the characteristics of local conservation laws, 
no local Hamiltonian structures, no local symplectic structures, and no local Noether operators, and no 
local inverse Noether operators. It should be pointed out that 
to the best of our knowledge, results on nonexistence of {\em any} local Hamiltonian structures or local symplectic structures for PDEs in more than two independent variables were not encountered in the literature. 
Moreover, the method of proof that we used can be readily applied to a number of other PDEs. 

The rest of the article is organized as follows. In Section~\ref{pre} we set the notation and recall some basic definitions required for the rest of the text, Section~\ref{mr} presents our main results whose proofs are then given in Section~\ref{pr}, and Section~\ref{cnd} contains conclusions and discussion. 


\section{Preliminaries}\label{pre}
Here we shall present the prerequisites required in order to state and prove our main results, mostly following \cite{olv93}; cf.\ also \cite{kvv}.  

We say that a function $f$ is {\em local} if it is smooth and depends at most on $x,y,t$
and finitely many of $u_{ij}=\partial^{i+j} u/\partial x^i\partial y^j$ where $i,j\in\{\ 0,1,2,\dots\}$.

From now on $x,y,t$ and $u_{ij}$ will be seen as coordinates on the appropriate jet space (or on the diffiety associated with \eqref{eqn} in the terminology of \cite{kvv}).

Let 
\begin{equation}\label{td}
D_x=\partial/\partial x+\sum\limits_{i,j=0}^\infty u_{i+1,j}\partial/\partial u_{ij},\ 
D_y=\partial/\partial y+\sum\limits_{i,j=0}^\infty u_{i,j+1}\partial/\partial u_{ij},\ D_t=\partial/\partial t+\sum\limits_{i,j=0}^\infty D_x^{i}D_y^{j}(F)\partial/\partial u_{ij},
\end{equation}
denote the operators of total derivatives adapted to equation (\ref{eqn}).

A {\em local conservation law} for \eqref{eqn} is an identity of the form
\begin{equation}\label{cld}
D_t(\rho)+D_x(\sigma)+D_y(\zeta)=0    
\end{equation}
where $\rho,\sigma,\zeta$ are local functions, not all of which are zero.

Then $\rho$ is called the {\em density} of the conservation law under study, and $\sigma$ and $\zeta$ are known as $x$- and $y$- {\em flux components}.

Let $\delta/\delta u$ denote the  operator of variational derivative
on local functions
\[\delta/\delta u=\sum_{i,j=0}^\infty (-1)^{i+j}D_x^iD_y^j \circ\partial/\partial u_{ij} \]
Note that for any local function $g$ the expression $\delta g/\delta u$ contains only finitely many terms, so there are no convergence issues.

For a local conservation law (\ref{cld}) its {\em characteristic} is defined as $\delta\rho/\delta u$ (in our setting this definition is readily seen to be equivalent to the more standard one, cf. \cite{kvv, olv93}). 

Let
\begin{equation}\label{cl2}D_t(\tilde\rho)+D_x(\tilde\sigma)+D_y(\tilde\zeta)=0  \end{equation}
be another local conservation law for 
\eqref{eqn}.

Quite obviously, a linear combination of \eqref{cld} and \eqref{cl2}, namely
\[
D_t(c_1\rho+c_2\Tilde{\rho})+D_x(c_1\sigma+c_2\tilde{\sigma})+D_y(c_1\zeta+c_2\tilde{\zeta})=0
\]
where $c_1$ and $c_2$ be constants, is again a local conservation law for \eqref{eqn}, i.e., local conservation laws for \eqref{eqn} form a vector space. 

A local conservation law for \eqref{eqn} is {\em trivial}, cf.\ e.g.\ Chapter 4 of \cite{olv93}, if there exist local functions $\alpha,\beta,\gamma$ such that
\[
\rho=D_x(\alpha)-D_y(\beta),\quad\sigma=D_y(\gamma)-D_t(\alpha),\quad \zeta=D_t(\beta)-D_x(\gamma)
\]

Two local conservation laws for \eqref{eqn} are {\em equivalent} if their difference is a trivial local conservation law.

For any local function $h$ define its linearization, or formal Frechet derivative, as \cite{olv93}
\[
D_h=\sum\limits_{i,j=0}^\infty \partial h/\partial u_{ij}D_x^iD_y^j
\]
(note that since $h$ is local, the above sum is actually finite, so there are no convergence issues). 

In particular, we have 
\[
D_F=-D_x^4-a D_y D_x-D_x\circ (f_u+f_{u_x}D_x), D_F^*=-D_x^4-a D_y D_x+D_x\circ (f_u D_x - D_x\circ f_{u_x})
\]
denote the linearization of the right-hand side of \eqref{eqn} and its formal adjoint (cf. below for the latter).

By definition $G$ is a {\em characteristic of local generalized symmetry} for \eqref{eqn} if $G$ is a local function that satisfies 
\begin{equation}
	\label{sym}
	D_t (G)-D_F(G)=0
\end{equation}
and $\gamma$ is a {\em local cosymmetry} for \eqref{eqn} if it is a local function that satisfies
\begin{equation}
	\label{cosym}
	D_t (\gamma)+D_F^*(\gamma)=0
\end{equation} 

It can be easily shown that for any local conservation law (\ref{cld}) its characteristic is necessarily a cosymmetry for (\ref{eqn}) but the converse in general is not true, i.e., there could be local cosymmetries that are not characteristics of local conservation laws. 

Note that 
there is no loss of generality in assuming local cosymmetries, characteristics of local generalized symmetries, densities and flux components of local conservation laws, etc.\ for \eqref{eqn} to not involve $t$-derivatives of $u$ or mixed derivatives thereof involving $t$, cf.\ e.g.\ \cite{olv93}.

For an operator of the form \[
L=\sum\limits_{i=0}^k \sum\limits_{j=0}^l h_{ij}D_x^i D_y^j, \] 
where $b_{ij}$ are local functions, assuming that $h_{kl}\neq 0$ introduce the obvious notation $k=\deg_x L$ and $l=\deg_y L$, with the standard convention, cf.\ e.g.\ \cite{olv93}, that $\deg_x0=\deg_y 0=-\infty$.

For example, for $F$ given by the right-hand side of \eqref{eqn} $\deg_x D_F=4$ and $\deg_y D_F=1$.\par


Here and below $\circ$ denotes composition of operators in total derivatives and for the above $L$ we define
\[D_t(L)=\sum\limits_{i=0}^k \sum\limits_{j=0}^l D_t(h_{ij})D_x^i D_y^j,\]
while the formal adjoint $L^*$ for the above $L$ is defined as
\[
L^*=\sum\limits_{i=0}^k \sum\limits_{j=0}^l (-D_x)^i (-D_y)^j \circ h_{ij}.
\]

\begin{defin}[cf. \cite{o}]\label{ino-def} An operator of the form
\[
N=\sum\limits_{i=0}^r \sum\limits_{j=0}^s h_{ij}D_x^i D_y^j 
\] 
where $h_{ij}$ are local functions,  is called 
a {\em local Noether operator} for \eqref{eqn}, resp.\ a 
{\em 
local inverse Noether operator} for \eqref{eqn},  if
\[
D_t (N)-D_F \circ N-N\circ D_F^*=0,
\]
resp.\ if 
\[D_t(N)+D_F\circ N+N\circ D_F=0\]
\end{defin}

The significance of these kinds of operators stems from the fact \cite{o}
that such operators map symmetries to cosymmetries or the other way around. More precisely, if $P$ is a local Noether operator for (\ref{eqn}), then for any local cosymmetry  $\gamma$ of (\ref{eqn}) we have that $P(\gamma)$ is a characteristic of local generalized symmetry for (\ref{eqn}). Likewise, if $J$ is a local inverse Noether operator for (\ref{eqn}), then for any characteristic $G$ of a local generalized symmetry for (\ref{eqn}) the quantity $J(G)$ is a local cosymmetry for \eqref{eqn}.\looseness=-1

While any local symplectic operator for (\ref{eqn}) is \cite{o} 
automatically is a local inverse Noether operator, the converse, 
generally speaking, is not true; we refer the reader to \cite{kvv, o} and references therein for further details on symplectic operators.

Likewise, while any local Hamiltonian operator for (\ref{eqn}) automatically is \cite{o} a local Noether operator for (\ref{eqn}), the converse in general does not hold; see e.g.\ \cite{kvv, olv93, s08} and references therein for further details on Hamiltonian operators.

Note that, just as for the recursion operators, cf.\ e.g.\ \cite{kvv, mas, o, olv77, olv93, s08} and references therein for those, (inverse) Noether, Hamiltonian and symplectic operators for nonlinear PDEs are often nonlocal, see e.g.\ \cite{kvv, olv93, vv}, but in the present paper we concentrate on the local case 
to avoid dealing with complicated issues of correct definition of action of such operators and passing from formal series in spirit of \cite{mik87, mik09, mikya, olv93} to actual operators, especially since we have more than two independent variables in (\ref{eqn}); see, however, Remark~\ref{rem-ino}.


\section{Main results}\label{mr}

We start with the following result readily proved by straightforward computation 
\begin{pro}\label{gen}For any smooth $f(u,u_x)$ equation \eqref{eqn} has infinitely many nontrivial conservation laws of the form
\begin{equation}\label{cly}
D_t(M u)+D_x((u_{xxx}+a u_y+f)M)=0	
\end{equation}
where $M$ is an arbitrary smooth function of $y$.
\end{pro}

Moreover, 
in certain special cases we have additional nontrivial local conservation laws:

\begin{pro}\label{extra}
In addition to the conservation laws from Proposition~\ref{gen},
Equation \eqref{eqn} with $a\neq 0$ and nonlinear smooth $f=f(u,u_x)$ 
%
further admits nontrivial local conservation laws 
not equivalent to those from Proposition~\ref{gen} if and only if 
$f$ is linear in $u_x$ and one of the following holds:

\medskip

\noindent{}i) there exist a smooth nonlinear function $g=g(u)$ of $u$ and constants $k_0$ and $k_1$ such that $f=g(u) u_x+k_1 u+k_0$; 


\medskip

\noindent{}ii) there exist a smooth nonlinear function $h=h(u)$ of $u$ and constants $c_0$ and $c_1$ such that $c_1\neq 0$ and $f=(c_1 \partial h(u)/\partial u+c_0)u_x+h(u)$

The additional nontrivial local conservation laws in both cases i) and ii) have the form 
\begin{equation}\label{clu}
		D_t(\zeta u)+D_x\left(-(u_{xx}-K_1 u_x+q)\zeta_x+(u_{xxx}+a u_y+f-K_2)\zeta\right)+D_y(-a u\zeta_x)=0,	
	\end{equation}
where for the case i) we have that $K_1=0$, $K_2=-k_0$, $q=q(u)$ is a smooth function of $u$ such that $\partial q(u)/\partial u=g(u)$, 
and
\begin{equation}\label{case-i}
\zeta=t (a \partial L/\partial y-k_1 L)+x L,
\end{equation}
where $L$ is an arbitrary smooth function of $y$,\\
while for the case ii) we have $q(u)=c_1 h(u)+(c_0+1/c_1^2)u$, $K_1=1/c_1$, $K_2=0$, and
\begin{equation}\label{case-ii}
\zeta=\exp(x/c_1+t(c_0/c_1^2+1/c_1^4))F(a t+c_1 y),
\end{equation}
where $F$ is an arbitrary smooth function of its argument. 
\end{pro}

It should be pointed out that even though $q(u)$ in the case i) is defined up to the addition of an arbitrary constant, say, $K_0$, the constant in question shows up in (\ref{clu}) only in the term $D_x(-K_0\zeta_x)$ which vanishes since in the case under study $\zeta$ is given by (\ref{case-i}). 

\begin{remark}
 Note that the constant $k_0$ in the case i) is not really essential, as $f$ in (\ref{eqn}) stands under the total $x$-derivative which annihilates the constant in question.
 
 On the other hand, the constant $c_0$ in the case ii) can be removed upon using the following change of variables: pass from $x$ to $X=x-c_0 y/a$ while keeping all other independent and dependent variables intact.   
\end{remark}

\begin{remark} 
It is readily checked that both cases of Proposition~\ref{extra} when additional conservation laws exist can also be presented in a more uniform 
fashion as follows using a somewhat different notation: 
there exist a smooth nonlinear function $g=g(u)$ of $u$ and constants $\tilde c_0$, $\tilde c_1$  and $\tilde c_2$ such that $f=u_x\partial g/\partial u+\tilde c_1 g +\tilde c_0 u+\tilde c_2$.  

The 
(additional) nontrivial local conservation laws from Proposition~\ref{extra} then still have the form (\ref{clu}), 
where now for $\tilde c_1=0$ 
\begin{equation*}
\zeta=x L+t(a\partial L/\partial y-\tilde c_0 L)
	\end{equation*}
where $L$ is an arbitrary smooth function of $y$, and for $\tilde c_1\neq 0$
\begin{equation*}
 \zeta=\exp(\tilde c_1 x+(\tilde c_0-\tilde c_1^3)y/a)F(y/\tilde c_1+a  t),
\end{equation*}
where $F$ is an arbitrary smooth function of its argument. 
	\end{remark}

When combined, Propositions~\ref{gen} and \ref{extra} provide a complete description of all cases when \eqref{eqn} with $a\neq 0$ and nonlinear $f$ admits nontrivial local conservation laws, and give explicit formulas for the conservation laws in question.

In particular, we have the following
\begin{cor}The only nontrivial local conservation laws admitted by the original Infeld--Rowlands equation, 
obtained from \eqref{eqn} upon setting $a=1$ and $f=u_x^2$, are those from Proposition~\ref{gen}.
\end{cor}

We also have two results concerning the cosymmetries; note that the first of those is stated separately as it does not assume the nonlinearity of the equation in question

\begin{pro}\label{cos} All local cosymmetries of equation \eqref{eqn} 
with $a\neq 0$ 
can depend at most on $x,y$ and $t$.
\end{pro}

\begin{pro}\label{co}
	The only local cosymmetries admitted by \eqref{eqn} with $a\neq 0$ and nonlinear $f$ are characteristics of local conservation laws listed in propositions~\ref{gen} and \ref{extra}.
\end{pro}

Moreover, we have 

\begin{pro}\label{hs}
Equation~\eqref{eqn} 
admits no nontrivial local Noether and inverse Noether operators. 
\end{pro}

As we have already mentioned in Section~\ref{pre} any local symplectic operator for (\ref{eqn}) is necessarily 
a local inverse Noether operator and a local Hamiltonian operator for (\ref{eqn}) is 
necessarily a local Noether operator, so we can immediately establish 
nonexistence of local Hamiltonian and local symplectic structures for (\ref{eqn}).\looseness=-1

\begin{cor}
Equation~\eqref{eqn} 
admits no nontrivial local Hamiltonian and symplectic operators.
\end{cor}

\begin{remark}\label{rem-ino}
In fact, using the method of proof of Proposition \ref{hs}, it is possible to show that \eqref{eqn} admits no nontrivial 
local Noether and inverse Noether
operators that can be represented as formal 
series of 
the form 
\[
\sum\limits_{i=-\infty}^r \sum\limits_{j=-\infty}^s b_{ij}D_x^i D_y^j 
\] 	
where $r$ and $s$ are any integers and $b_{ij}$ are local functions. 	
\end{remark}

\section{Proofs of the main results}\label{pr}



%


\medskip

\noindent{\em Proof of Proposition~\ref{extra}} By Proposition~\ref{co}, which will be proved later, the only cosymmetries admitted by \eqref{eqn}, are

\noindent 1) $M(y)$, where $M$ is an arbitrary smooth function of $y$, for any smooth $f$,

\noindent 2) $\zeta$ given by \eqref{case-i} or \eqref{case-ii} if $f$ is nonlinear and satisfies extra assumptions from case i) or ii) from  
Proposition~\ref{extra}. 

Now 
recall that in our setup a characteristic of a local conservation law is necessarily a local cosymmetry (the other way around this does not hold in general).

By the above, all cosymmetries of (\ref{eqn}) depend at most on $x,y,t$, and it is immediate that to any such cosymmetry $\chi=\chi(x,y,t)$ there corresponds, up to the addition of a trivial local conservation law, a local conservation law for (\ref{eqn}) with the density $\rho= \chi(x,y,t)u$, and the result readily follows upon compting the associated flux components for the appropriate densities $\rho$.
$\Box$

\medskip

\noindent{\em Proof of Proposition~\ref{cos}} 
Suppose that $\gamma$ is a local cosymmetry for \eqref{eqn}, and let $k=\deg_x D_\gamma$ and $l=\deg_y D_\gamma$. 

It is immediate that to prove that $\gamma$ depends at most on $x,y,t$ is equivalent to proving that $D_\gamma=0$.

Seeking a contradiction, assume that $D_\gamma\neq 0$, 
%
Then obviously $k\geqslant 0$, and upon repeated use of (\ref{td}) we find that taking the partial derivative of (\ref{cosym}) w.r.t.\ $u_{k+4,l}$ 
yields
\[
2\partial\gamma/\partial u_{kl}=0, 
\]
Taking the above into account and acting by the operator of partial derivative w.r.t.\ $u_{k+4,l-1}$ on (\ref{cosym}) now yields
\[
2\partial\gamma/\partial u_{k,l-1}=0, 
\]
and continuing in the same fashion we find that
for all $j=0,1,\dots,l$ 
\[
\partial\gamma/\partial u_{kj}=0, 
\]
hence we see, using the definition of $D_\gamma$, that in fact $\deg_x D_\gamma$ is at most $k-1$, which contradicts our initial assumption $\deg_x D_\gamma=k$. 
The only way to resolve this contradiction is to assume that $D_\gamma=0$, and hence $\gamma$ can depend at most on $x,y,t$
$\Box$

\medskip

\noindent{\em Proof of Proposition~\ref{co}} First of all, by Proposition~\ref{cos}, 
any local cosymmetry of \eqref{eqn} can depend at most on $x,y,t$. 

Then the condition \eqref{cosym} boils down to
\begin{equation}\label{cosym-1}
\diff{\gamma}{t}-\diff{^4\gamma}{x^4}-a\diff{^2\gamma}{x \partial y} - f_{u_x}\diff{^2\gamma}{x^2} + f_u\diff{\gamma}{x} - f_{u_xu_x}u_{xx}\diff{\gamma}{x} - f_{uu_x}u_x\diff{\gamma}{x} = 0,
\end{equation}

Differentiating (\ref{cosym-1}) w.r.t.\ $u_{xx}$ yields
\[f_{u_xu_x}\diff{\gamma}{x}=0.\]
We readily see that if $f_{u_xu_x}\neq 0$, i.e., $f$ is not linear in $u_x$, $\gamma$ must be independent of $x$. 
Now, if $\gamma$ is independent of $x$, \eqref{cosym-1} becomes 
$\partial\gamma/\partial t=0$
and we conclude that $\gamma$ is also independent of $t$. Hence the only possible cosymmetry in this case is an arbitrary function $M(y)$.

Let us now assume 
the function $f$ to be linear in $u_x$, i.e., to have the form
$$f = f_1 u_x + f_0,$$
where $f_i$ are 
smooth functions of $u$ only. 

The equation \eqref{cosym-1} then takes the following form: 
\begin{equation}\label{cosym-f-lin_u_x}
\diff{\gamma}{t}-\diff{^4\gamma}{x^4}-a\diff{^2\gamma}{x \partial y} - f_1\diff{^2\gamma}{x^2} + \frac{\partial f_0}{\partial u}\diff{\gamma}{x} = 0.
\end{equation}

The structure of solutions to this equation 
depends on whether $\partial f_0/\partial u, f_1$ and 1, considered as functions of $u$, are linearly dependent or not, so we split the analysis of \eqref{cosym-f-lin_u_x} into several cases. 

\medskip

\noindent{\em Case 1:} $f_1$ is linearly independent from $\partial f_0/\partial u$ and 1.

Then in order for \eqref{cosym-f-lin_u_x} to hold we must, in particular, require that the coefficient at $f_1$ vanishes, and thus $\partial^2\gamma/\partial x^2=0$ so
\[\gamma=\gamma_0+x\gamma_1\]
where $\gamma_i$ are smooth functions of $y$ and $t$.

Substituting this expression for $\gamma$ back into \eqref{cosym-f-lin_u_x} yields
\begin{equation}\label{cosym-f-lin_u_x-1}
x\diff{\gamma_1}{t}+\diff{\gamma_0}{t}-a\diff{\gamma_1}{ y} + \frac{\partial f_0}{\partial u}\gamma_{1} = 0,
\end{equation}
whence, upon equating to zero the coefficient at $x$, we immediately see that $\gamma_1$ in fact depends on $y$ alone.

The rest of (\ref{cosym-f-lin_u_x-1}) then yields
\begin{equation}\label{cosym-f-lin_u_x-1.next}
\diff{\gamma_0}{t}-a\diff{\gamma_1}{ y} + \frac{\partial f_0}{\partial u}\gamma_{1} = 0.
\end{equation}

We now have two subcases.

\medskip 

\noindent{\em Subcase 1a:} $\partial f_0/\partial u$ is linearly independent from 1. 

Then for \eqref{cosym-f-lin_u_x-1.next}  to hold we must, in particular, equate 
to zero the coefficient at $\partial f_0/\partial u$ in \eqref{cosym-f-lin_u_x-1.next}. Upon doing so 
we have again, just as above, that $\gamma_1=0$. Thus 
$\gamma$ has to be independent of $x$ and consequently, by virtue of (\ref{cosym-f-lin_u_x-1.next}), of $t$. and hence the only possible cosymmetry in this case is again an arbitrary function of $y$ only, $M(y)$. Thus, if $\partial f_0/\partial u$ is linearly independent from 1 while $f_1$ is linearly independent from both $\partial f_0/\partial u$ and 1, there are no local cosymmetries for \eqref{eqn} other than those being characteristics of conservation laws from Proposition \ref{gen}.


\newpage
\noindent{\em Subcase 1b:} $\partial f_0/\partial u$ is linearly dependent from 1, 
so there are constants $k_0$ and $k_1$ such that $f_0=k_1 u+k_0$. 

Then \eqref{cosym-f-lin_u_x-1.next} boils down to
\[\diff{\gamma_0}{t}-a\diff{\gamma_1}{ y} + k_1 \gamma_{1} = 0,\]

As $\gamma_1$ per above is independent of $t$, we find that the general solution of the above equation reads
\[
\gamma_0=M+t\left(a \diff{\gamma_1}{ y} - k_1 \gamma_{1}\right)
\]
where $M$ is an arbitrary smooth function of $y$ and so is $\gamma_1$.

It is now clear that $M$ again corresponds to the cosymmetry being the characteristic of the conservation law from Proposition~\ref{gen} while the arbitrary function $\gamma_1$ which gives rise to an additional set of cosymmetries
\begin{equation}\label{cs-i}
x L+t\left(a \diff{L}{ y} - k_1 L\right)
\end{equation}
where we have for convenience relabelled the arbitrary function by $L$ instead of $\gamma_1$.  

It is readily checked that cosymmetries (\ref{cs-i}) are precisely characteristics (\ref{case-i}) of the conservation laws from case i) of Proposition~\ref{extra}.

\medskip

\noindent{\em Case 2:} 
$f_1$ is linearly dependent from 1 and $\partial f_0/\partial u$. 

First of all observe that then, as \eqref{eqn} is nonlinear by assumption, the functions 1 and $\partial f_0/\partial u$ must be linearly independent as functions of $u$, otherwise $f$ would be linear in both $u$ and $u_x$ and so \eqref{eqn} would be linear as well.

Thus, we now have
\begin{equation}
 f_1=c_1\partial f_0/\partial u+c_0   
\end{equation}
where $c_i$ are arbitrary constants and $\partial f_0/\partial u$ is nonconstant.


Since 1 and $\partial f_0/\partial u$ must be linearly independent as per the above, we get from \eqref{cosym-f-lin_u_x} upon separately equating to zero the coefficients at 1 and  $\partial f_0/\partial u$  that 
\begin{equation}\label{gt}
    \diff{\gamma}{t}-\diff{^4\gamma}{x^4}-a\diff{^2\gamma}{x \partial y} - c_0\diff{^2\gamma}{x^2}= 0.
\end{equation}
and
\begin{equation}\label{gx}
    \frac{\partial\gamma}{\partial x}=c_1 \diff{^2\gamma}{x^2}.
\end{equation}

Let us split this into two subcases : $c_1=0$ and $c_1 \neq 0$

\medskip

\noindent{\em Subcase 2a:} 
$c_1=0$.
Then $\gamma$ is independent of $x$ by \eqref{gx} and then also independent of $t$ by (\ref{gt}). Thus, if $c_1=0$ then $\gamma$ is an arbitrary smooth function of $y$ alone and the only possible local cosymmetries in this case are again 
those being characteristics of conservation laws from Proposition \ref{gen}.

\medskip

\noindent{\em Subcase 2b:} $c_1\neq 0$. Then $\partial^2\gamma/\partial x^2=(1/c_1)\partial\gamma/\partial x$, and (\ref{gt}) boils down to
\begin{equation}\label{gt1}
    \diff{\gamma}{t}-\frac{1}{c_1^3}\diff{\gamma}{x}-a\diff{^2\gamma}{x \partial y} -\frac{c_0}{c_1}\diff{\gamma}{x}= 0.
\end{equation}

General smooth solution of \eqref{gx} is obvious:
\[
\gamma=\gamma_0+\gamma_1\exp(x/c_1)
\]
where $\gamma_i$ are arbitrary smooth functions of $y$ and $t$.

Substituting this into (\ref{gt1}) and equating to zero separately the coefficients  at the two linearly independent functions of $x$, 1 and $\exp(x/c_1)$, yields
\[\partial\gamma_0/\partial t=0\]
so $\gamma_0$ in fact depends on $y$ alone, and 
\[
\frac{\partial \gamma_1}{\partial t}-\frac{a}{c_1} \frac{\partial\gamma_1}{\partial y}-\frac{c_0 c_1^2+1}{c_1^4}\gamma_1=0.
\]
The general smooth solution of the above equation reads 
\[
\gamma_1=\exp\left(\frac{(c_1^2 c_0+1)t}{c_1^4}\right) F\left(a t+c_1 y\right)
\]
where $F$ is an arbitrary smooth function of its argument.

It is clear that $\gamma=\gamma_0(y)$ for smooth functions $\gamma_0(y)$ are the characteristics for the conservation laws from Proposition~\ref{gen} while
\[
\exp\left(\frac{x}{c_1}+\frac{(c_1^2 c_0+1)t}{c_1^4}\right) F\left(a t+c_1 y\right)
\]
with arbitrary smooth $F$ are  precisely characteristics (\ref{case-ii}) of conservation laws from case ii) of Proposition \ref{extra}, which concludes the analysis of the subcase in question and thus of Case 2.

Summing up all the above special cases, we see that indeed all nontrivial local cosymmetries for \eqref{eqn} with $a\neq 0$ and nonlinear $f=f(u,u_x)$ are characteristics of nontrivial local conservation laws listed in Propositions \ref{gen} and \ref{extra}. $\Box$ 

We now proceed to Proposition~\ref{hs}.

\noindent{\em Proof of Proposition~\ref{hs}}. 
Let $P$ of the form 
\begin{equation}\label{ho}
P=\sum\limits_{i=0}^r \sum\limits_{j=0}^s p_{ij}D_x^i D_y^j 
\end{equation} 
where $P_{ij}$ are local functions, be a local Noether operator for \eqref{eqn}, i.e.,  $\tilde P=0$, where
\[
\tilde P=D_t(P)-D_F^*\circ P-P\circ D_F
\]
We readily see that the leading term of $\tilde P$ is $-2p_{rs} D_x^{r+4}D_y^s$ and since we require that $\tilde P=0$, this leading term must vanish, i.e., $p_{rs}=0$, and moreover $p_{rj}=0$ for all $j=0,\dots,s-1$.

Continuing by replacing $r$ by $r-1$ in the above considerations and so on establishes that $p_{ij}=0$ for all $i$ and $j$, so $P=0$, i.e.\eqref{eqn} admits no nontrivial Noether operators of the form \eqref{ho}, which completes the part of the proof concerning the Noether operators.

Likewise, assume that 
\[
B=\sum\limits_{i=0}^k \sum\limits_{j=0}^l b_{ij}D_x^i D_y^j, \] 
where $b_{ij}$ are local functions, is a local inverse Noether operator for (\ref{eqn}), i.e.,  it satisfies $\tilde B=0$ where
\[
\tilde B=D_t(B)+D_F^* \circ B+B\circ D_F.
\]
In a similar fashion as for the Noether operator case we see that $b_{kj}=0$ for all $j=0,\dots,l$ and then that $b_{ij}=0$ for all $i$ and $j$, so $B=0$ and the result follows.
$\Box$

%
%

\section{Conclusions and discussion}\label{cnd}

In the previous sections we gave a complete characterisation of all cases when the generalized Infeld--Rowlands equation \eqref{eqn} with $a\neq 0$ and nonlinear $f$ admits nontrivial local conservation laws and have listed all such inequivalent local conservation laws of all orders modulo the addition of trivial ones.

It turned out in particular that equation (\ref{eqn}) for any $a$ and $f$ admits an infinite set of nontrivial local conservation laws parameterized by an arbitrary function of $y$ as described in Proposition~\ref{gen}.

If $f$ is nonlinear then, as shown in Proposition~\ref{extra}, 
nontrivial local conservation laws beyond those from Proposition~\ref{gen} can exist only provided $f$ is linear in $u_x$ and moreover has a special form given in case i) or ii) of Proposition~\ref{extra}.


The conservation laws that we found have many potential applications including e.g.\ in the numerical simulation for \eqref{eqn}, cf.\ the discussion 
in Section~\ref{in}. 

In this connection we also point out that the conservation laws from Proposition~\ref{gen} because of their special form have one more possible application. Namely, using these, it is possible to introduce a nonlocal potential, say $w$, for (\ref{eqn}) defined by the formulas
\begin{equation}\label{nonlo}
w_x=u,\quad w_t=-(u_{xxx}+a u_y+f) 
\end{equation}
and 
it could be of interest to study nonlocal symmetries, nonlocal conservation laws etc.\ involving this potential; this is, however, beyond the scope of the present paper whose focus is on local objects and structures 
(cf.\ e.g.\ \cite{kvv} and references therein for a general introduction to nonlocal objects in the geometric theory of PDEs).

Furthermore, we have shown that the only local cosymmetries admitted by \eqref{eqn} with nonlinear $f$ are the characteristics of the above conservation laws from Propositions~\ref{gen} and \ref{extra}.

Note that it could be of interest to  study symmetries of (\ref{eqn}) with nonlinear $f$ and, in particular, to find out whether the cases from Proposition~\ref{extra} when (\ref{eqn}) admits additional conservation laws are distinguished in some fashion from the symmetry point of view as well. 

Finally, we have shown that (\ref{eqn}) 
admits no  nontrivial local Noether operators and local inverse Noether operators and hence a fortiori no nontrivial local Hamiltonian and symplectic structures. While similar results are known for the case of evolution equations in two independent variables, see e.g.\ \cite{hs, mik87} and references therein, to the best of our knowledge this is the first result of the kind for the case of more than two independent variables, and the 
same method can be readily applied to establish analogous results for many other equations and systems, like e.g.\ the degenerate Burgers equation from \cite{vps}. 

\subsection*{Acknowledgements}

This research was supported by the Specific Research grant SGS/13/2020 of 
Silesian University in Opava.

I would like to thank Artur Sergyeyev for stimulating discussions and valuable comments. 

A significant part of the computations in the paper was performed using the computer algebra package
\emph{Jets} \cite{jets} whose use is hereby gratefully acknowledged.
 

\end{document}